\documentclass[11pt,letter]{svproc}

\usepackage[left=1in, right=1in, top=1in,	bottom=1in]{geometry}

\usepackage{here}
\usepackage{microtype}%if unwanted, comment out or use option "draft"
\usepackage{amsmath,mathtools,etoolbox}
\usepackage{mathrsfs}
\usepackage{xcolor}
\usepackage{marvosym}
\usepackage{array,multirow,makecell,graphicx,tabulary}
\usepackage[textwidth=0.8in,textsize=scriptsize,backgroundcolor=yellow]{todonotes}
\usepackage{tikz}
\usetikzlibrary{arrows,snakes,backgrounds,patterns,matrix,shapes,fit,calc,shadows,plotmarks}
\providecommand{\os}{\overline{S}}
\providecommand{\slam}{\mathcal{S}_{\textit{Lam}}}
\providecommand{\suni}{\mathcal{S}_{\textit{Uni}}}
\providecommand{\sopt}{\mathcal{S}_{\textit{Opt}}}
\renewcommand{\sf}{s_{\textit{f}}}
\renewcommand{\sb}{s_{\textit{b}}}
\providecommand{\sbi}{s_{\textit{bi}}}

\newcommand{\ce}[1]{\lceil #1 \rceil}

\newcommand{\ol}[1]{\overline{#1}}
\newcommand{\udl}[1]{\underline{#1}}

\newcommand{\Ll}{\mathscr{L}}
\newcommand{\Ul}{\mathscr{U}}

\newcommand{\zl}{\udl{z}}
\newcommand{\zu}{\ol{z}}

\let\theparentequation\theequation
\patchcmd{\theparentequation}{equation}{parentequation}{}{}

\renewenvironment{subequations}[1][]{%              optional argument: label-name for (first) parent equation
  \refstepcounter{equation}%
  \setcounter{parentequation}{\value{equation}}%    parentequation = equation
  \setcounter{equation}{0}%                         (sub)equation  = 0
  \def\theequation{\theparentequation\alph{equation}}%
  \let\parentlabel\label%                           Evade sanitation performed by amsmath
  \ifx\\#1\\\relax\else\label{#1}\fi%               #1 given: \label{#1}, otherwise: nothing
  \ignorespaces
}{%
  \setcounter{equation}{\value{parentequation}}%    equation = subequation
  \ignorespacesafterend
}

\newcommand*{\nextParentEquation}[1][]{%            optional argument: label-name for (first) parent equation
  \refstepcounter{parentequation}%                  parentequation++
  \setcounter{equation}{0}%                         equation = 0
  \ifx\\#1\\\relax\else\parentlabel{#1}\fi%         #1 given: \label{#1}, otherwise: nothing
}

\begin{document}
\title{Optimum Search Schemes \\for Approximate String Matching Using Bidirectional FM-Index\vspace*{-2ex}}

\author{Kiavash Kianfar\inst{1,*} \and
Christopher Pockrandt\inst{2,3}\and
Bahman Torkamandi \inst{1} \and \\
Haochen Luo \inst{1}\and
Knut Reinert \inst{2,3,**}}
\authorrunning{Kianfar et al.} % abbreviated author list (for running head)
\institute{Department of Industrial and Systems Engineering, Texas A\&M University \\ *\email{kianfar@tamu.edu}
\and
Department of Computer Science and Mathematics, Freie Universit\"at Berlin, Germany\\
\and
Max Planck Institute for Molecular Genetics, Berlin, Germany\\ **\email{knut.reinert@fu-berlin.de}\vspace*{-2ex}
}
%
%\author[1]{Kiavash Kianfar}
%% \thanks{Texas A\&M University
 %%   (\protect\email{kianfar@tamu.edu}, \protect\url{http://ise.tamu.edu/people/faculty/kianfar/modal/index.html}).}
 %\author[2]% \and
  %Christopher Pockrandt\thanks{FU Berlin
    %(\protect\email{christopher.pockrandt@fu-berlin.de}, \protect\url{http://reinert-lab.de}).}
  %\and
  %Bahman Torkamandi \thanks{Texas A\&M University (\protect\email{bahman\_1988@tamu.edu}).}
  %\and
  %Haochen Luo \thanks{Texas A\&M University (\protect\email{hcluo@tamu.edu}).}
  %\and
  %Knut Reinert\thanks{FU Berlin (\protect\email{knut.reinert@fu-berlin.de})}
%}
%\institute{Department of Computer Science and Mathematics, Freie Universit\"at Berlin, Germany
%\and
%Department of Industrial and Systems Engineering, Texas A\&M University
%}

\pagestyle{plain}

\maketitle\vspace*{-3ex}

\begin{abstract}
%\todoki{Dear Knut, this is how far Bahman and I have gotten in revising the manuscript. The idea is the same I explained in my email last week. We took out FAMOUS name to get the attention to what we have done in the paper, we motivate development of famous at the end and state it as ongoing future work. Bahman did a pass on it and now I am going over. I am done up to the end of section 4. We have left some comments as well. We would appreciate your and Chris' feedback.}

Finding approximate occurrences of a pattern in a text using a full-text index is a central problem in bioinformatics and has been extensively researched. \emph{Bidirectional} indices have opened new possibilities in this regard allowing the search to start from anywhere within the pattern and extend in both directions. In particular, use of \emph{search schemes} (partitioning the pattern and searching the pieces in certain orders with given bounds on errors) can yield significant speed-ups. However, finding  \emph{optimal search schemes} is a difficult combinatorial optimization problem.

\parindent=2em Here for the first time, we propose a mixed integer program (MIP) capable to solve this optimization problem for Hamming distance with given number of pieces. Our experiments show that the optimal search schemes found by our MIP significantly improve the performance of search in bidirectional FM-index upon previous ad-hoc solutions. For example, approximate matching of 101-bp Illumina reads (with two errors) becomes $35$ times faster than standard backtracking. Moreover, despite being performed purely in the index, the running time of search using our optimal schemes (for up to two errors) is comparable to the best state-of-the-art aligners, which benefit from combining search in index with in-text verification using dynamic programming. As a result, we anticipate a full-fledged aligner that employs an intelligent combination of search in the bidirectional FM-index using our optimal search schemes and in-text verification using dynamic programming that will outperform today's best aligners. The development of such an aligner, called FAMOUS (Fast Approximate string Matching using OptimUm search Schemes), is ongoing as our future work.

\vspace*{-4ex}
\end{abstract}

\small
\begin{keywords}
  FM index, bidirectional, read mapping, approximate matching, mixed integer programming, optimization\vspace*{-2ex}
\end{keywords}

\normalsize
\section{Introduction}\vspace*{-2ex}
\parskip=0.05in
% describe importance of approximate search and give
% overview of possible solutions leading to approximate seeds in (unidirectional-)indices.
Finding approximate occurrences of a string in a large text is a fundamental  problem in computer science with numerous applications.  The \emph{approximate string matching (ASM) problem} for Hamming distance considered in this paper is defined as follows: Given a number of mismatches $K$, a string (here referred to as a \emph{read}) of length $R$, and a text of length $T$, composed of characters from {an} alphabet of size $\sigma$, find a substring of the text whose Hamming distance to the read is at most $K$. A similar definition can be provided for ASM for edit distance where in addition to mismatches, insertions and deletions, are also considered.

Solving the ASM problem has become
especially important in bioinformatics due to the advances in sequencing technology during the last years. The mainstream second generation sequencing techniques like Illumina produce reads of length $150$-$250$ with an error rate of about $1\%$, mostly substitutions caused by the sequencing technology.
%\todoki{The $1\%$ you mention is caused by sequencing technology, correct? Knut: yes}.
Other sequencing technologies, e.g., Pacific Bioscience or Oxford Nanopore, produce much longer reads but with a higher error rate (in the range of $15\%$) containing both substitutions and insertions/deletions. A standard problem is to map the reads back to a reference genome while taking into account the errors introduced by the sequencing technology as well as those caused by biological variation, such as SNPs or small structural variations. Such a problem is almost always modeled as the ASM problem for Hamming or edit distance.

There are two main algorithmic strategies to address the ASM problem for large input sizes (in number of reads and size of the text): \emph{filtering} and \emph{indexing}.
%Filtering approaches quickly exclude large regions of the reference where no approximate match can be found. This can, for example, be done by identifying short regions in the reference (also known as $k$-mer) that share a short piece of the read without errors, often called a \emph{seed}. Regions that do not share such a short region are filtered out. In addition to seeding filters, there are also filters based on shared $q$-gram counts \cite{Burkhardt1999} or based on the pigeonhole lemma.
%The second main idea is to preprocess the reference sequence, the set of reads, or both, in a more intricate way.
In this work, we focus on the indexing approach. Here, the main idea is to preprocess the reference sequence, the set of reads, or both, in a more intricate way.
Such preprocessing into \emph{full-text string indices} has the benefit that we usually do not have to scan the whole reference, but can conduct queries much faster at the expense of larger memory consumption. String indices that are currently used are the suffix array \cite{Manber:1990wh} and enhanced suffix array \cite{Abouelhoda:2004hw}, and affix arrays \cite{maass2003linear,strothmann2007affix}, as well as the FM-index \cite{Ferragina:2000vl}, a data structure based on the Burrows-Wheeler Transform (BWT) \cite{burrows1994block} and some auxiliary tables. For an in-depth discussion see \cite{Reinert:2015ds}.
Such indices are used to compute exact matches between a query and a text as a subroutine in backtracking approaches. For the ASM problem for Hamming or edit distance, the existing algorithms all have exponential complexity in $K$ (e.g. \cite{Karkkainen2007,Vroland:2016ca}), and are hence only suited for small $K$.

 %\violet{Affix trees are a generalization of suffix trees using the duality of suffix trees caused by the suffix links. Although they are capable of performing a bidirectional search by incorporating the suffix tree and the suffix tree of the reverse text \cite{maass2003linear}, they suffer from large memory consumption resulting in invention of new data structure, called affix array similar to affix tree in functionality however with lower memory consumption \cite{strothmann2007affix}. }
% first use was by Lam
Lam et al. \cite{LamLiTam09} introduced \emph{bidirectional} FM indices to speed up ASM for Hamming distance.
%\violet{To distinguish between different types of strings in this paper, we refer to the input string that is to be approximately found in (matched to) the given large text as the \emph{read} (for example, the read can be a DNA sequencing read and the large text can be a reference genome). }
For the cases $K=1$ and $2$, they partitioned  the read  into $K + 1$ equal pieces, and argued  that performing approximate matching on a certain combination of these pieces in a bidirectional index amounts to \emph{faster} approximate matching of the whole read. This combination is such that all possible \emph{mismatch patterns}, i.e., all possible distributions of $K$ mismatches among the pieces, are covered.  The main idea behind improved speed is that a bidirectional index not only can start the search from the beginning (or end) of the read, but also from the beginning (or end) of any of the pieces. Therefore, we can start the search from a middle piece and then expand it to the left or right into adjacent pieces in any order we like.  By choosing multiple appropriate orderings of pieces for this purpose, we can perform a much faster ASM compared to a unidirectional search because we can enforce exact or near-exact searches on the first pieces in the partition, significantly reducing the number of backtrackings, while using different orderings of pieces to ensure all possible mismatch patterns are still covered.

\setcounter{page}{1}
% describe Kucherovs idea
Kucherov et al. \cite{KucSalTsu16}  formalized and generalized this idea by defining  the concept of \emph{search schemes}. Assume a read can be partitioned into a given number of pieces, denoted by $P$ (not necessarily equal to $K+1$). The pieces are indexed from left to right. A search scheme $\mathcal{S}=\{(\pi_s,L_s,U_s), s=1,\ldots,S\}$ is a collection of $S$ searches, where each search $s$ is designated by a triplet $(\pi_s,L_s,U_s)$. $\pi_s$ is a permutation of $1,\ldots,P$ and denotes the order in which the pieces of the partition are searched in search $s$. If $\pi_{s,i}=j$, then piece $j$
%\todo{I am not very happy with the $s_i$, its more $\pi_s(i)=j)$. I guess then we would also have to change $L_{s_i}$ to $L_s(i)$ }
%\todoki{I changed the notation to $\pi_{s,i}$. And made the same changes for L and U. I think this makes the notation fine. Knut: Yes}
is searched at {position $i$} in the order (shortly referred to as \emph{iteration} $i$ in this paper). Due to the way a bidirectional index works, the permutation $\pi_s$ must satisfy the so-called \emph{connectivity} condition, i.e, a piece $j$ can appear at iteration $i>1$ in the permutation only if at least one of pieces $j-1$ or $j+1$ have appeared at an iteration before $i$. $L_s$ and $U_s$ each are strings of $P$ numbers. $L_{s,i}$,  is the lower bound on the cumulative number of mismatches allowed at iteration $i$ of search $s$, and $U_{s,i}$ is the upper bound on this value. %\todoki{Knut, I would rather we do not refer to the figure at this stage. Would that be a problem?} %In Figure \ref{fig:trie} this formalism is exemplified on the same small example used by Kucherov. The figure shows the original Lam search scheme consisting of three searches, the standard backtracking which is a search scheme consisting of a single search in this scheme, and our optimal solution.

Having this formal framework, the answer to the \emph{Optimal Search Scheme} problem, defined as follows, can potentially have a great impact on improving the running time of ASM using a Bidirectional index (ASM-B).

\noindent \emph{{\bf Optimal Search Scheme Problem:} What is the search scheme that minimizes the number of steps in ASM-B while ensuring all possible mismatch patterns are covered?}

It turns out that this is a very difficult combinatorial optimization problem  due to several reasons: There are a large number of attributes that define a solution (including $S$, $P$, size of each piece, and $(\pi_s,L_s,U_s)$ for each search) with a large number of possibilities for each attribute; the solution must satisfy complex combinatorial constraints; and, calculating the objective function, i.e., number of steps in the ASM-B algorithm, for a given solution is complicated.

Kucherov et al. \cite{KucSalTsu16} presented some interesting results contributing initial insight into this key problem. More specifically, they assumed the number of steps in the ASM-B algorithm with a given search scheme, is a constant factor of the (weighted) total number of substrings enumerated by the algorithm in all searches. Assuming that a randomly generated read is to be matched to a randomly generated text, they presented a method to calculate this objective function for a given search scheme.
They then showed that unequal pieces in the partition can potentially improve the objective function compared to equal pieces, and presented a dynamic programming (DP) algorithm that for a single prespecified search, with given $P$ and $(\pi,L,U)$, finds the optimal sizes of pieces assuming that we only calculate the objective function as the total number of substrings up to a limited length (justified by total randomness of the read and the text); see \cite{KucSalTsu16} for more details. In fact, the superiority of this DP  over explicit enumeration is only due to this assumption.
Nevertheless,  this DP is very inefficient, and most importantly, it only finds the optimal piece sizes for a \emph{prespecified} search. In other words, it does not address the problem of finding an optimal search scheme which calls for determining $S$ and all attributes of each search in the search scheme, and ensuring that they cover all mismatch patterns.
%\todo{Kiavash I took this part out to save space since i think it is not that important here. If you like. put it back in, or we could replace it by a short sentence "They also presented some other interesting results, but none of them addressed the problem of computing optimal search schemes}

Kucherov et al. \cite{KucSalTsu16} also presented solutions for another limited problem, i.e., lexicographically minimizing the lexicographically maximal $U$ string (critical $U$ string) in a search scheme, only for $P=K+1$ or $K+2$ and assuming that the $L$ strings for all searches contain only zeros. The usefulness of these solutions is justified by the high probability that the search with the critical $U$ string has the largest share in the objective function; see  \cite{KucSalTsu16} for details. Again from the perspective of finding an optimal search scheme, this result has similar limitations. Only one of the attributes ($U$) of one of the searches for two specific values of $P$ are optimized by fixing all $L$ strings, which is far from designing a globally optimal search scheme as defined above. Consequently, in their computational experiments, Kucherov et al. \cite{KucSalTsu16} use a greedy algorithm based on this limited result to construct search schemes with unknown quality and only optimize the piece sizes for these schemes using their DP.

% describe our contributions
In this paper, for the first time, we propose a method to solve the optimal search scheme problem for ASM-B with Hamming distance, for any given $P$ and equal-size pieces. Our method is based on a novel and powerful mixed integer linear program (MIP) that gets $K$, $R$, $P$, and an upper bound on $S$, denoted by $\os$, as input, and provides, as its solution, all the attributes of the exact optimal search scheme (MIP methodology for optimization has been addressed in many references such as \cite{NemWol88,Wol98}).  We present our MIP and provide the results of our computational study on the characteristics of its optimal solution and its running time, for different values of its input parameters.
% we will extend the work of \viogray{Kucherov et al.}{Kucherov et al.\cite{KucSalTsu16}} in several respects. \viogray{First}{Firstly}, we slightly adapt the definitions and constraints of search schemes (\viogray{which led to the small mistake in the Lam scheme}{which has led to a small mistake in Lam's scheme}). \viogray{secondly we give for the first time}{ Secondly, for the first time we propose} a method of computing an optimal search scheme \viogray{for}{with respect to} Hamming distance \viogray{and fixed partition length}{for equal length partitions} by means of an integer linear program.

{Next we conduct an experiment using a bidirectional search (for Hamming and edit distance) that performs the search based on the optimal search schemes we have obtained from our MIP. This bidirectional index search is implemented in SeqAn \cite{Reinert:2017gm} and uses a recent fast implementation of bidirectional indices \cite{Pockrandt:2017hc} based on EPR dictionaries.  }
We show that, for practical ranges of various input parameters, the number of substrings for the optimal search schemes found by our MIP can reduce to as small as half the number of substrings in the unidirectional complete backtracking.
%\viogray{We show that with our optimal search schemes, FAMOUS improves the search time for approximate string matching dramatically. Searching
%for \emph{all} occurrences of
%% \todoki{Are they simulated? Looked to me you used real reads from an accession number.
%% also added ``on human genome sequence''}
%Illumina reads on human genome sequence using FAMOUS is up to $3$, $14$, and $35$ times faster than unidirectional search for $K=1$, $2$, and $3$ Hamming distance errors, respectively.}
{In another experiment, we search for all occurrences of Illumina reads in the human genome completely in the bidirectional index using our optimal search schemes and show that our time is is up to $3$, $14$, and $35$ times faster than the standard backtracking search in the index for $K=1$, $2$, and $3$ Hamming distance errors, respectively. }%\todoki{A comparison with Lam scheme for the case P=3 and K=2 may be a good addition to the computational experiments}
Although our MIP finds the optimal search schemes for Hamming distance, when we used its optimal schemes with the edit distance, we got similar improvements.

%\viogray{Additionally,
%%we evaluated FAMOUS using Illumina reads and the human genome sequence for Hamming and edit distance
%we perform two other experiments using FAMOUS: In the first one, we search for \emph{all} approximate occurrences of Illumina reads in the genome (all-mapping), and in the second one, instead of {all-mapping}, we do the search according to the typical \emph{strata}-based strategy, i.e., we search for all best occurrences plus a number of errors up to a global error upper bound.
%We show that FAMOUS in strata mode is already competitive with Yara \cite{Yara}. The all-mapping results suggest that a combination of our optimal search schemes and a verification strategy might lead to a tool outperforming all current ASM approaches for genomic reads.
%We note that the optimal searches found by our MIP are implemented in the SeqAn library for efficient sequence analysis, and hence, can in the future speed up \emph{any} tool part of which includes searching for approximate occurrences of a read.}{}

The drastic improvement over standard backtracking gained by using our optimal search schemes for bidirectional search in index suggests that the performance of read mappers that utilize an index can be significantly improved. To gauge this potential, we even challenged our optimal search schemes by performing a pure index-based search using them and comparing the performance with the full-fledged state-of-the-art aligners that benefit from using a combination of search in index and verification in text using dynamic programming. We noticed that our optimal schemes are so effective that although they are employed in the pure index-based search, the resulting times for $K=1$, $2$, and $3$, are competitive with the state-of-the-art aligners which use a combination of index search and verification in text, in strata mode. In the all-mapping mode, the results are better for $K=1$,  and competitive for $K = 2$. These observations suggest that a full-fledged aligner that employs an intelligent combination of search in the bidirectional FM-index using our optimal search schemes and verification in text using dynamic programming can outperform today's best approximate read mappers. The development of such an aligner, called FAMOUS (Fast Approximate string Matching using OptimUm search Schemes) is ongoing in our group as future work.

%\todoki{For me this sentence implies that these optimal schemes are already being used in your algorithms which are available online for everyone to download. Are we at that stage yet? Or should we say will be implemented?}
%\todo{Other contributions? Don't think so, but I leave the todo for later†}

We will introduce our MIP for solving the optimal search scheme problem in Section \ref{sec:oss} and discuss our computational studies on solving this MIP. Then in Section \ref{sec:comp}, we present the computational impact of using the optimal search schemes obtained from the MIP on solving ASM-B using various realistic scenarios. In Section \ref{sec:famous}, we will motivate the potential impact of our optimal search schemes in developing a full-fledged read mapper and mention FAMOUS as our future work. We will conclude in Section \ref{sec:con} by summarizing our contributions and raising several open problems and possible extensions.\vspace*{-2ex}

\section{Solving Optimal Search Scheme Problem using MIP} \vspace*{-2ex}
\label{sec:oss}
{In this section, we provide our MIP-based methodology for finding optimal search schemes after presenting some preliminaries. We will then follow with a brief computational report on solving our MIP in order to find the optimal search schemes, including its optimal objective value as a function of its input parameters, its solution running time, and its convergence rate to optimal solution). }

Our MIP is for Hamming distance, but as mentioned before, based on our computational experiments (Section \ref{sec:famous}), its optimal schemes for Hamming distance are very good (but not necessarily optimal) search schemes for the edit distance as well. \vspace*{-3ex} %As a result, although our discussion in this section is focused on Hamming distance, in our computational experiments (Section \ref{sec:famous}), we will use the Hamming distance optimal schemes for the Edit distance problem as well.
\subsection{Preliminaries}\vspace*{-1.2ex}
\label{sec:pre}
Our MIP presented in Section \ref{sec:mip} will solve the optimal search scheme problem assuming $P$ is given as an input (is not a decision variable in optimization) and all $P$ pieces of the partition are equal in length, i.e., $R=mP$, where $m$ denotes the length of any piece. Note that these assumptions pose no practical restrictions. Solving problems which include $P$ as a decision variable and allow unequal pieces is part of our future research plan. Given the upper bound on Hamming distance $K$ (maximum number of mismatches) as an input, a \textit{mismatch pattern} is a particular distribution of $h$ mismatches among the $P$ pieces, for any $h\le K$. Specifically, the {mismatch pattern} $q$ is a string of $P$ integers $a_{q,1}\ldots a_{q,P}$ such that $a_{q,j} \in \{0,\ldots,\min\{m,K\}\}$ for $j=1,\ldots,P$, and $\sum_{j=1}^P a_{q,j} = h$. For given $K$ and $P$, we denote the set of all possible mismatch patterns by $\mathcal{M}$. Note that if $K\le m$ then $|\mathcal{M}| = \sum_{h=0}^K \binom{h+P-1}{h}$.
Given a search $s=(\pi_s,L_s, U_s)$, a mismatch pattern $q$ is said to be \textit{covered} by $s$ if at every iteration $i=1,\ldots,P$ of $s$, $L_{s,i} \leq \sum_{t=1}^i a_{q,\pi_{s,t}} \leq U_{s,i}$, i.e., the cumulative number of mismatches up to iteration $i$ is between the allowed lower and upper bounds of search $s$. A search scheme $\mathcal{S}$ is feasible if and only if every mismatch pattern in $\mathcal{M}$ is covered by at least one search in $\mathcal{S}$.

\begin{figure}[t]
\centering
\begin{minipage}{2.18in}
\centering
\frame{\includegraphics[width=2.18in]{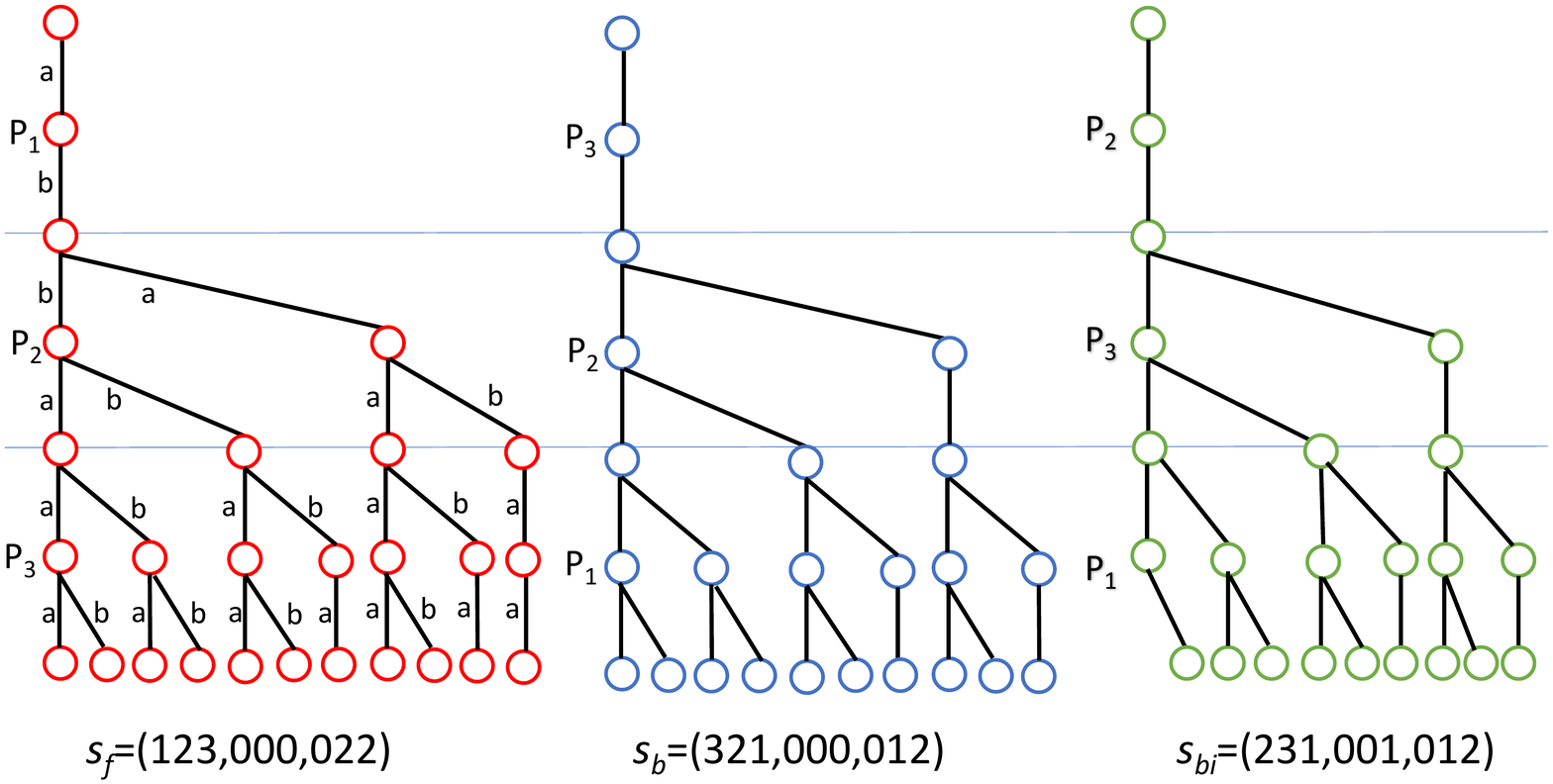}}%\hspace{0.5in}

\small(a)
\end{minipage}%\hspace{0.2in}
\begin{minipage}{2.18in}
\centering
\frame{\includegraphics[width=2.18in]{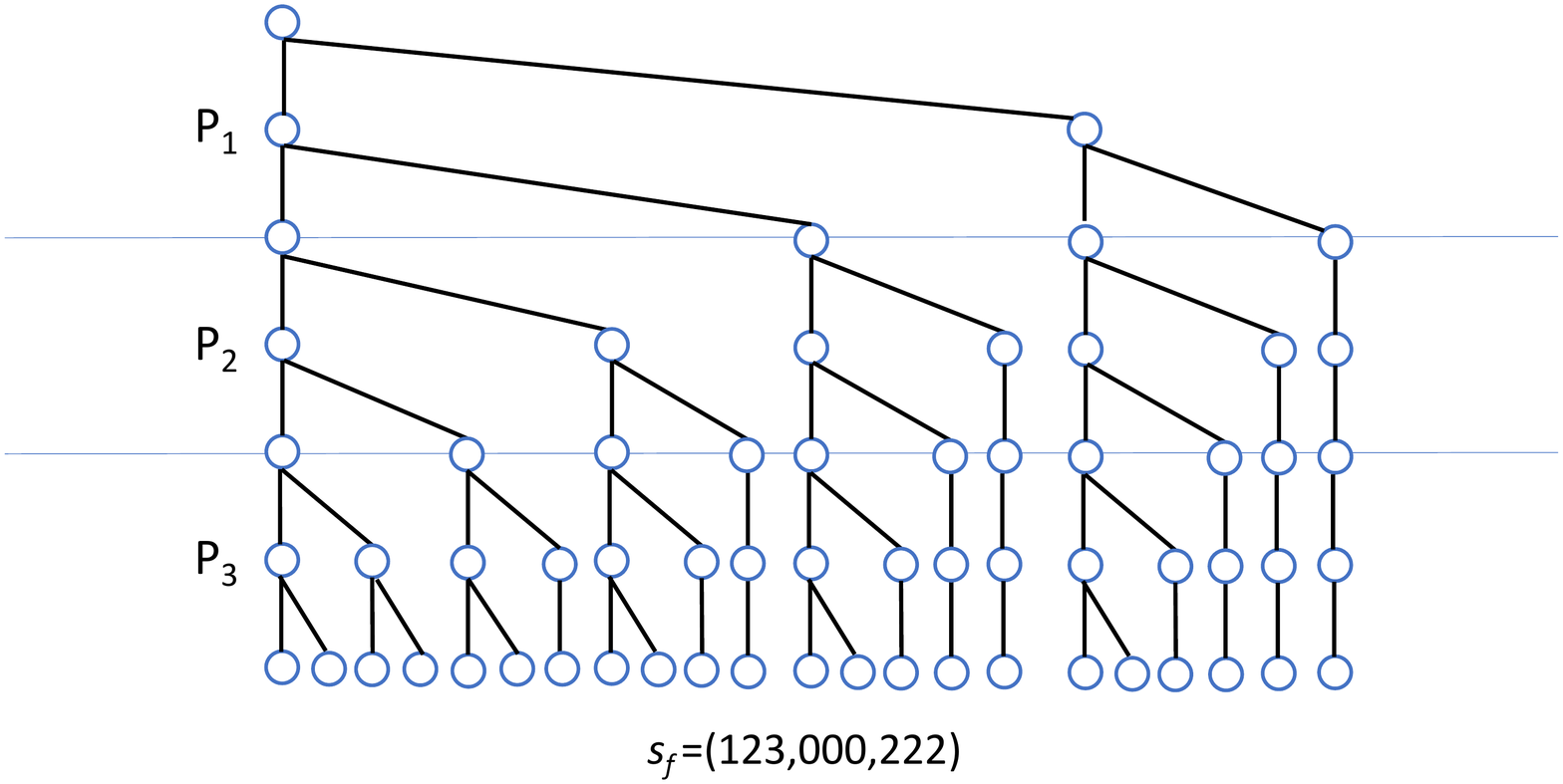}}%\hspace{0.5in}

\small(b)
\end{minipage}%\hspace{0.2in}
\begin{minipage}{2.18in}
\centering
\frame{\includegraphics[width=2.18in]{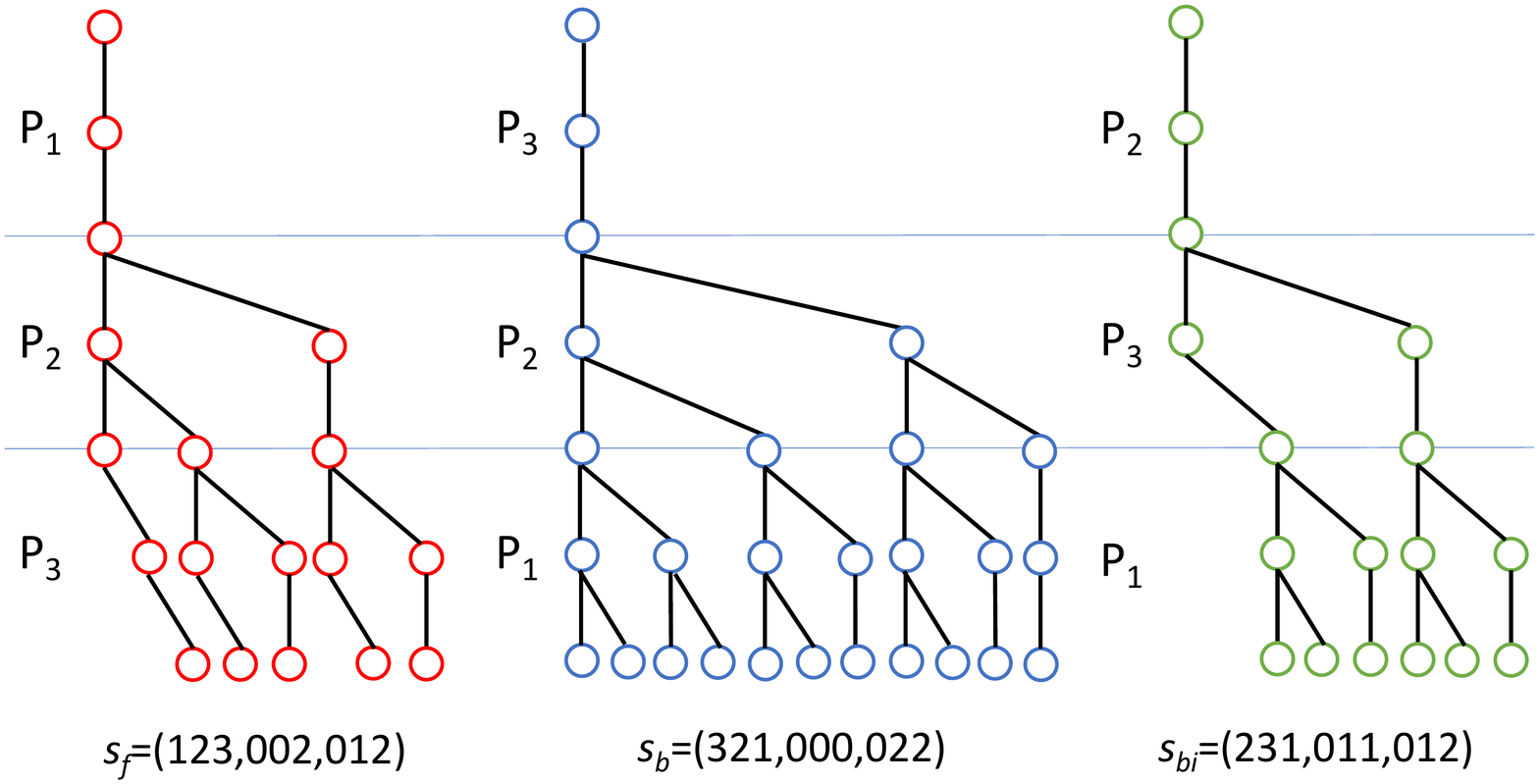}}%\hspace{0.5in}

\small(c)
\end{minipage}

\caption{{\bf (a)} The search of Lam et al. \cite{LamLiTam09} as described by Kucherov \cite{KucSalTsu16} %\cite{LamLiTam09}
 for $K=2$ and  $P=3$, i.e., $\slam=\{\sf=(123,000,022),\sb=(321,000,012),\sbi=(231,001,012)\}$,  shown for the  read ``abbaaa'' from the alphabet $\{\text{a,b}\}$, i.e., $R=6$ and $\sigma=2$. The read is partitioned into $P_1=$ab, $P_2=$ba, and $P_3=$aa.  Partition borders are shown by horizontal lines.  A vertical and a diagonal edge represent a match and a mismatch, respectively. Edge labels are only shown for $s_f$ for a cleaner picture. The search corresponding to each trie is designated underneath it by its $(\pi,L,U)$. The number of edges in $\slam$ tries is  71.  %It can be seen that all possible error configurations are covered by at least one of the three searches.
{\bf (b)} The unidirectional search scheme $\suni=\{\sf=(123,000,222)\}$ for the same problem. The number of edges in $\suni$ is 62, i.e., for this particular problem, in which $R$ is very small, $\slam$ enumerates even more substrings than $\suni$ (if all possible substrings are present in the text). Of course, if $R$ gets larger, the situation is reversed,  making $\slam$ more efficient than $\suni$ as reported by  Lam et al. \cite{LamLiTam09}.
{\bf (c)} The optimal search scheme $\sopt=\{\sf=(123,002,012),\sb=(321,000,022),\sbi=(231,011,012)\}$ for the same problem, found by our MIP.  The total number of edges in $\sopt$ (optimal number of edges) is 59, which is less than that of $\suni$, and significantly less than that of $\slam$. As shown in Section \ref{sec:famous}, for bigger problems, the reduction in the total number of edges of the optimal search scheme found by our MIP compared to the unidirectional search is much more significant (up to 50\%). \vspace*{-0.25in}
} \label{fig:trie}
\end{figure}

A search scheme can be visualized by representing each of its searches as a trie that captures all substrings enumerated by the search. Each edge at a level of the trie corresponds to a character of the alphabet at that level of search.  A vertical edge represents a match, and a diagonal edge represents a mismatch. Fig. \ref{fig:trie}(a) shows the tries associated with the search scheme presented by Lam et al. \cite{LamLiTam09} for $K=2$ and $P=3$, $\slam$, applied on  the six-character read ``abbaaa'' from alphabet $\{a, b\}$ (note that the tries are slightly different from the ones given in \cite{KucSalTsu16}, which {contained} a small error). %We omitted the edge description for the two other tries for clarity).
Fig. \ref{fig:trie}(b) shows a search scheme with a single unidirectional search (complete backtracking), $\suni$, for the same problem, and Fig.  \ref{fig:trie}(c) shows the optimal search scheme, $\sopt$, found by our MIP, for the same problem. Each one of the three schemes in Fig. \ref{fig:trie} covers all $10$ mismatch patterns, namely $\{000,001,010,100,011,101,110,002,020,200\}$. Interestingly, the three searches $s_f,s_b,s_{bi}$ in  $\sopt$ cover the mismatch patterns $\{002,011\}$, $\{000,010,100,110,020,200\}$, and $\{001,101\}$, respectively, which is indeed a partition of all mismatch patterns (see open problems in Section \ref{sec:con}), whereas in $\slam$, the searches $s_f$ and $s_b$ both cover $000$ and $010$ redundantly.

Following the method of Kucherov et al. \cite{KucSalTsu16}, we define the performance of a search scheme as the number of forward and backward steps taken by the ASM-B algorithm, which is equal to the total number of substrings enumerated by all searches in the scheme. We assume a single step of forward or backward search in the bidirectional index takes the same amount of time. The tries of any search scheme in Fig. \ref{fig:trie} contain all possible substrings of length $R$. The number of substrings in each trie is equal to the number of edges (or total number of non-root nodes). If the text contains all substrings of length $R$, the search enumerates all substrings in the tries; hence, the performance of the search scheme can be measured by the total number of edges in the search scheme. Otherwise, only a subset of the substrings in the tries will be enumerated depending on whether they occur in the text or not.
To address the performance measure in this latter case, Kucherov et al. \cite{KucSalTsu16} assumed the read and the text are randomly and independently drawn from the alphabet according to a uniform distribution, and hence, calculated the expected number of substrings enumerated by the scheme as the sum, over all non-root nodes of the tries, of the probability that the corresponding substring appears in the text. As a result, they presented a \emph{weighted} sum of number of edges as the measure of performance. Due to the assumption of complete randomness and independence of the read and the text, they show that the weights of the edges at levels lower than $\ce{log_{\sigma}T}+c_{\sigma}$
%\todoki{T is the lenght of the text defined in the first paragraph of the paper}
of the tries, where $c_{\sigma}$ is that $((\sigma-1)/\sigma)^{c_{\sigma}}$ is sufficiently small, are almost zero meaning that they can be dropped from the weighted summation.

	%\viogray{If edit distance is considered, a similar trie can be constructed as depicted in Figure \ref{fig:lam_scheme_edit}}{ Furthermore, we propose a new trie in Figure \ref{fig:lam_scheme_edit}  to accomodate for edit distance in case of considering edit distance } for the bidirectional (green) search \viogray{from}{depicted in} Figure \ref{fig:lam_scheme}.

For the main application of our interest, i.e. ASM of DNA sequence reads to reference genomes, the assumption of randomness and independence of the read and the text is far from reality. Calculating the expected number of substrings enumerated by a scheme calls for significant more study on determining probabilities that DNA sequence reads of particular length from a sample occur in the reference genomes. As currently there is no trivial answer to this problem, in this paper, we use the same performance measure of total number of edges in the tries of the search scheme even for the case where not all substrings occur in the text. Of course, our MIP can be easily modified to incorporate any other weighting scenario which might be proposed in the future.

%According
Adapting the method from \cite{KucSalTsu16}, the total number of edges in the search scheme is calculated by \vspace*{-1ex}
\begin{equation}\small
\label{eq:total}
\sum\nolimits_{s=1}^S \sum\nolimits_{l=1}^R \sum\nolimits_{d=0}^K n_{s,l,d},\vspace*{-0.5ex}
\end{equation}
where $n_{s,l,d}$ is defined as the number of edges at level $l$ of the trie of search $s$ that end at nodes corresponding to substrings with $d$ cumulative mismatches up to that level. The value of $n_{s,l,d}$ can be calculated using the following recursive equation, which is an adaptation of the formula in \cite{KucSalTsu16}: \vspace*{-1ex}
\begin{equation}\small
\label{eq:nsld}
n_{s,l,d} =
n_{s,l-1,d}+(\sigma-1) n_{s,l-1,d-1} \text{\qquad\qquad for  } l \geq 1 \text{   and  } \Ll_{s,l} \leq d \leq \Ul_{s,l},\vspace*{-1ex}
\end{equation}
where, by definition, $n_{s,0,0}=1$, $n_{s,0,-1}=0$ and $n_{s,0,d} = 0$, for $d \ge 1$, $s=1,\ldots,S$, and $\Ll_{s,l}$ and $\Ul_{s,l}$ denote
the smallest and largest cumulative number of mismatches that can occur at level $l$ of the trie of search $s$, respectively, calculated as $ \Ll_{s,l}=\max\{L_{s,{\ce{l/m}-1}},L_{s,{\ce{l/m}}}- m\ce{l/m} + l \}$ and $\Ul_{s,l}=\min\{U_{s,{\ce{l/m}}},\Ul_{s,l-1}+1\}$. Here $\ce{l/m}$, the smallest integer greater than or equal to $l/m$, would be the index of the iteration in which level $l$ falls, and by definition, $\Ll_{s,0}=\Ul_{s,0}=0$, for $s=1,\ldots,S$. For example, for  search $\sbi$ of $\sopt$, we have $\Ll_{\sbi}=(0,0,0,1,1,1)$ and $\Ul_{\sbi}=(0,0,1,1,2,2)$. \vspace*{-2ex}
%\todoki{Bahman, do we need this sentence here?} \violet{In fact, for pieces of equal length, our MIP is capable of implementing equation \eqref{eq:nsld} without employing $\Ll_{s,l}$ and $\Ul_{s,l}$ through constraints \eqref{na}, \eqref{nb}, \eqref{ia}, and \eqref{ib}. \vspace*{-2ex}}

%\begin{lstlisting}[caption={Useless code},label=list:8-6,captionpos=t,float,abovecaptionskip=-\medskipamount]
%for i:=maxint to 0 do
%begin
%    j:=square(root(i));
%end;
%\end{lstlisting}
%

\subsection{MIP Formulation of Optimal Search Scheme Problem }\vspace*{-1ex}
\label{sec:mip}

Our MIP formulation, presented below, solves the optimal search scheme problem assuming $P$ is given as an input and pieces are all equal in length. More specifically, for given $K$, $R$, $P$, and $\os$, this MIP finds the  search scheme with minimum total number of edges among all feasible search schemes that have at most $\os$ searches. The optimal solution to the MIP provides the $(\pi,L,U)$ of all searches in the optimal search scheme. The objective value of this optimal solution provides the minimum total number of edges (substrings) achievable among all feasible search schemes. \vspace*{-1ex}

\allowdisplaybreaks

\begin{equation}\small
\min \hspace{1em} \sum\nolimits_{s=1}^{\os} \sum\nolimits_{l=1}^R \sum\nolimits^K_{\color{black}{d=0}}  n_{s,l,d} \label{obj}
\end{equation}
subject to
\begin{subequations}\small
\begin{align}
&\sum\nolimits_{i=1}^{P} x_{s,i,j} =1 \hspace{-20em} & \text{for all $s$ and $j$} \label{oa} \\
&\sum\nolimits_{j=1}^{P} x_{s,i,j} =1 \hspace{-10em} & \text{for all $s$ and $i$} \label{ob}  \\[1em]
%&&\nonumber\\
\nextParentEquation
&\sum\nolimits_{h=1}^i x_{s,h,j} -\sum\nolimits_{h=1}^i x_{s,h,j-1} = t^+_{s,i,j} - t^-_{s,i,j} \hspace{-20em} & \begin{multlined}[t][1.5in] \text{for all $s$, $i=2,\ldots,P-1$,}\\ \text{$j=1,\ldots,P+1$}\end{multlined} \label{ca} \\
&\sum\nolimits_{j=1}^{P+1} (t^+_{s,i,j} + t^-_{s,i,j}) = 2 \hspace{-10em} &\text{for all $s$, $i=2,\ldots,P-1$} \label{cb}\\[1em]
%&&\nonumber\\
\nextParentEquation
&d  -(L_{s,\lceil l/m \rceil}-m\lceil l/m \rceil + l)   +1 \le (R+1) \zl_{s,l,d} \hspace{-10em} & \text{for all $s$, $l$, and $d$} \label{na}\\
&U_{s,\lceil l/m \rceil}+1-d \le (K+1) \zu_{s,l,d} \hspace{-10em} &\text{for all $s$, $l$, and $d$}\label{nb} \\
& \scalebox{0.95}{${{l}\choose{d}}$}(\sigma-1)^d(\zu_{s,l,d}+\zl_{s,l,d}-2) \le n_{s,l,d}- n_{s,l-1,d}-(\sigma-1)n_{s,l-1,d-1} \hspace{-10em} &  \text{ for all $s$, $l$, and $d$} \label{nc} \\[1em]
%&&\nonumber\\
\nextParentEquation
& L_{s,i} \le L_{s,i+1} \hspace{-10em} &  \text{for all $s$, and $i=1,\ldots,P-1$} \label{ia} \\
& U_{s,i} \le U_{s,i+1} \hspace{-10em} & \text{for all $s$, and $i=1,\ldots,P-1$} \label{ib} \\[1em]
%&&\nonumber\\
\nextParentEquation
&  L_{s,i}+K(\lambda_{q,s}-1) \le \sum\nolimits_{h=1}^i\sum\nolimits_{j=1}^P a_{q,j}x_{s,h,j} \le U_{s,i} + K(1-\lambda_{q,s}) \hspace{-5em} & \text{for all $q$, $s$, and $i$}  \label{qa} \\
& \sum\nolimits_{s=1}^{\os} \lambda_{q,s} \ge 1 \hspace{-10em} &\text{for all $q$} \label{qb}\\[1em]
%&&\nonumber\\
\nextParentEquation
& n_{s,l,d} \ge 0 \hspace{-10em} &\text{for all $q$, $s$, $i$, $j$, $l$, and $d$} \label{sa} \\
& L_{s,i}, U_{s,i} \ge 0 \quad \text{Integer} \hspace{-10em} &\text{for all  $s$ and $i$} \label{sb} \\
& x_{s,i,j}, \lambda_{q,s}, \zu_{s,l,d}, \zl_{s,l,d}, t^+_{s,i,j}, t^-_{s,i,j} \in \{0,1\} \hspace{-10em} &\text{for all $q$, $s$, $i$, $j$, $l$, and $d$} \label{sc}
\end{align}
\end{subequations}

The objective function \eqref{obj} minimizes the total number of edges as calculated by \eqref{eq:total} with $n_{s,l,d}$ as defined before. The binary variables $x_{s,i,j}$ capture the assignment of pieces to iterations, i.e., $x_{s,i,j} = 1$ if piece $j$ is searched at iteration $i$ of search $s$, and $x_{s,i,j}=0$ otherwise. We define $x_{s,i,0}=x_{s,i,P+1}=0$ to simplify presentation of constraints. At optimality, these variables determine the $\pi_s$ values for the optimal search scheme.  Constraints \eqref{oa} and \eqref{ob} make sure that for any search $s$, only one piece is assigned to an iteration and only one iteration is assigned to a piece.

Constraints \eqref{ca}-\eqref{cb} ensure the connectivity of the pieces
%i.e., piece $j$ can be searched at iteration $i$ only if at least one of pieces $j-1$ or $j+1$ has been searched at an iteration before $i$. Constraints \eqref{ca}-\eqref{cb}
and are in fact linearization of the following constraint using auxiliary binary variables $t^+_{s,i,j}$ and $t^-_{s,i,j}$:\vspace*{-1.5ex}
\begin{equation}\small
\sum\nolimits_{j=1}^P \left\vert\sum\nolimits_{h=1}^i x_{s,h,j} -\sum\nolimits_{h=1}^i x_{s,h,j-1}\right\vert = 2 \qquad \text{for all $s$ and $i=2,\ldots,P-1$}, \label{eq:ex1}\vspace*{-1.5ex}
\end{equation}
which is one way to enforce connectivity of pieces. The term $\sum\nolimits_{h=1}^i x_{s,h,j}$ will have a binary value which denotes whether or not piece $j$ has been searched at any of iterations $1$ to $i$ of search $s$. The term $\sum\nolimits_{h=1}^i x_{s,h,j-1}$ captures the same notion for piece $j-1$. If at any iteration all searched pieces form a connected block on the read, the value of $\sum\nolimits_{h=1}^i x_{s,h,j} -\sum\nolimits_{h=1}^i x_{s,h,j-1}$ will be equal to 1 only for one $j$, $-1$ for another $j$, and $0$ for all other $j$'s, which is ensured by \eqref{eq:ex1}, and hence its linearization.
%\todo{I am through, except for checking the ILP and writing the intro paragraphs which i will do on the weekend. Very nice... The only notational problem i see is the $\pi_{s_i}$ and the different version of $L$ which can be confusing. Then again, to change all that would be major work. So i guess we leave it like that}
%\todoki{I fixed the remaining part as well. Fixed the issue with $\pi_{s_i}$ by changing the index usage for $\pi$,$L$ and $U$ all over. }

%Connectivity can be enforced alternatively by the following constraint:
%\begin{equation}
%x_{sij} \leq \sum_{h=1}^i x_{sh,j-1} + \sum_{h=1}^i x_{sh,j+1} \hspace{5em}\text{for all $s,i$ and $j$}. \label{ineq:alca}
%\end{equation}
%Intuitively, \eqref{ineq:alca} \viogray{means}{indicates} that partition $j$ can be searched only if partition $j-1$ or $j+1$ \viogray{is searched before}{was searched}. However, this is computationally less efficient \viogray{than the}{than} constraints \eqref{ca}-\eqref{cb}.
%\todo{Bahman: Drop the alternative constarint that we didn't use}

Constraints \eqref{na}-\eqref{nc} enforce calculation of $n_{s,l,d}$ based on the recursive equation \eqref{eq:nsld} with the help of binary variables $\zl_{s,l,d}$ and $\zu_{s,l,d}$. Due to \eqref{na}, if $d \ge L_{s,{\ce{l/m}}}- m\ce{l/m}  + l $, then $\zl_{s,l,d}=1$, and due to \eqref{nb}, if $d \leq U_{s,{\ce{l/m}}}$, then $\zu_{s,l,d}=1$. Calculation of equation (\ref{eq:nsld}) is then enforced by \eqref{nc}. When $\zl_{s,l,d}=\zu_{s,l,d}=1$, \eqref{nc} reduces to $n_{sld}- n_{s,l-1,d}-(\sigma-1)n_{s,l-1,d-1} \geq 0$, which implies $n_{sld}- n_{s,l-1,d}-(\sigma-1)n_{s,l-1,d-1} = 0$ since the objective function is to be minimized. If any of  $\zl_{s,l,d}$ or $\zu_{s,l,d}$ is equal to $0$, \eqref{nc} does not enforce anything as $-{l\choose d}(\sigma-1)^d$ is a lower bound on the right-hand side of \eqref{nc}.
Constraints \eqref{ia}-\eqref{ib} ensure $L_{s,i}$ and $U_{s,i}$ are non-decreasing as they are cumulative values. Constraints \eqref{qa}-\eqref{qb} ensure feasibility of the search scheme. $\lambda_{q,s}$ is a binary variable designating whether or not mismatch pattern $q$ is covered by search $s$. Constraint \eqref{qa} forces $\lambda_{q,s}=0$ if search $s$ does not cover mismatch pattern $q$ and constraint \eqref{qb} ensures every mismatch pattern $q$ is covered by at least one search, for $q=1,\ldots,|\mathcal{M}|$.

Constraints \eqref{oa}-\eqref{sc} are enough to formulate the MIP; however, we have noticed that imposing the additional constraints \vspace*{-1.5ex}
\begin{subequations}\small
\begin{align}
& x_{1PP} = 1  &  \label{va} \\
& \sum\nolimits_{t=s}^{\os} \sum\nolimits_{k=1}^{j-1} x_{t,1,k} \le (\os-s+1) (1- x_{s,1,j})\hspace{-1.2em} & \text{  for all $s$ and $j=2,\ldots,P$} \label{vb}
\end{align}
\end{subequations}\vspace*{-3.5ex}
\small\begin{align}
&\sum\nolimits_{j=1}^{P-i+1} x_{sij} + \sum\nolimits_{j=i}^{P} x_{sij} = 1 \hspace{4em} & \text{ for all $s$ and $i \ge \lceil P/2 \rceil +1$} \label{w}
\end{align}\normalsize
strengthens the formulation while preserving at least one optimal solution, resulting in faster solution time for the MIP. Constraints \eqref{va} and \eqref{vb} eliminate some symmetry in the solution space.
For every search scheme, there is an equivalent search scheme obtained by reversing all $\pi_s$, $s=1,\ldots,\os$. Constraint \eqref{va} eliminates one of these two equivalent solutions in each pair by forcing piece $P$ to be assigned to iteration $P$ in the first search, eliminating the solutions in which piece $1$ is assigned to iteration $P$.
For any search scheme, another equivalent search scheme can be obtained by permuting the indices of searches within the scheme. Existence of only one of the search schemes obtained by this index permutation in the feasible solution set is enough. This can be achieved by sorting (in ascending order) the searches based on the piece assigned to their first iteration. This is done by constraint \eqref{vb}, which does not allow pieces $1,\ldots,j-1$  to be assigned to the first iteration of searches $s,\ldots,\os$ if piece $j$ is assigned to the first iteration of search $s$.
In addition to symmetry elimination, notice that the connectivity condition of pieces implies that the piece assigned to iteration $P$  is either piece $1$ or piece $P$, and in general, the piece assigned to iteration $i\ge \ce{P/2}+1$ is one of pieces $1,\ldots, P-i+1, i,\ldots, P$. Constraint \eqref{w} enforces this property, which strengthens the formulation, and according to our computational tests, reduces the running time of the MIP.
\begin{remark}
A powerful feature of our MIP is that $\os$ is an upper bound on the number of searches, i.e., our MIP finds the optimal search scheme among all schemes with at most $\os$ searches. In our MIP, variables are defined for $\os$ searches, and if the optimal search scheme has $S^*<\os$ searches, our MIP generates $\os-S^*$ \emph{empty} searches, i.e. searches in which $L_{s,i} > U_{s,i}$ for some $i$. \qed \vspace*{-2ex}
% which implies the search covers no mismatch patterns.
 \end{remark}

\subsection{Solving MIP}
\label{sec:mipcomp}
\vspace{-2ex}
We used CPLEX 12.7.1 solver \cite{CPLEX127} to solve our MIP by implementing the code in C++ using CPLEX Callable Library. All instances were run over four 28-core nodes (2.4 GHz Intel Broadwell) with 64GB of memory per node. We ran our MIP solver for instances generated for a broad range of parameters $K$, $R$, $\os$, and $P$ and gave each instance a 3-hour time limit. Fig. \ref{fig:graph}(a) in the Appendix is a small representative of our results. It shows the optimal objective value (total number of edges) for $R=100$, $K=1,\ldots,4$, $P=5,6$, and $\os=1,\ldots,5$. If the problem is not solved to optimality in 3 hours, the best solution found within this time limit is shown. The optimal objective value does not show a consistent change pattern in terms of change in $P$; however, as expected, it increases as $K$ increases, as $R$ increases (not shown), and as $\os$ decreases.  In all instances, the optimal objective value shows a sharp drop from $\os=1$ to $\os=2$, then a modest drop to $\os=3$, and negligible change beyond $\os=3$, generating empty searches in many cases. Therefore, as long as $\os=5$, it is advisable to use $\os=3$ instead if we would like to reduce the MIP running time and still find an optimal or near-optimal solution for $\os=5$. We also noticed that the optimal search scheme obtained by our MIP is not sensitive to the value of $R$ (see open problems in Section \ref{sec:con}). Therefore, when $R$ is large, it is advisable to solve the MIP for a much smaller reasonable value of $R$, e.g., $R=KP$, in order to get a solution that is most probably optimal for the large $R$ in a much shorter amount of time. %\vspace*{-3ex}

%\vspace*{-0.2in}
%Using the MIP formulation, we were able to solve considerable size problems to optimality. For instance, we were able to solve a problem with $K=4$, $R=100$, $P=3$, and $\os=3$ to optimality in $5802$ seconds. However, more complicated cases reached the run time limit of $3$ hours without proving the optimality of the solution. Consequently, it is important to investigate the rate of convergence of the solutions found during execution of MIP to the  the optimal solution. Figure \ref{fig:optprog} \todo{fix} illustrates the ratio of the best solutions found by MIP during its execution to the final optimal objective value plotted against run time for some instances which reached optimality. It can be observed that in all cases within $0.1\%$ to $1\%$ of its total run time, the MIP finds a solution which is finally proved to be optimal or very close to the optimal after MIP execution is complete. This means that the MIP solver finds ched \todo{fix this} to vicinity of the optimal solution in $10$ seconds which is about $0.1 \% $ to $ 1 \% $ of time to optimality for various cases.
%

Using the MIP formulation, we were able to solve considerable size problems to optimality. For instance, we were able to solve a problem with $K=4$, $R=100$, $P=3$, and $\os=3$ to optimality in $5802$ seconds. However, more complicated cases reached the time limit of $3$ hours without proving solution optimality. Consequently, it is important to investigate the rate of convergence of the solutions found during execution of MIP to the optimal solution. Fig. \ref{fig:graph}(b) illustrates the ratio of the best solutions found by MIP during its execution to the final optimal objective value plotted against running time for some instances which reached optimality. We observe that in all cases, within $0.1\%$ to $1\%$ of the total running time, the MIP finds a solution which is finally proved to be optimal or very close to the optimal after MIP execution is complete. In other words, the MIP solver finds optimal or near optimal solutions very early on and spends the rest of its time ensuring that no better solution exists. This can be partly due to the remaining symmetry in the solution space. Nevertheless, from practical perspective, this is an attractive property because, when the input parameters are much larger, we can run the MIP for a short time and find solutions which are most probably optimal or near-optimal.\vspace{-3ex}

{ \section{Search-in-Index Computational Gains Achieved by Optimum Schemes}\vspace*{-2ex}
\label{sec:comp}
%\vspace{-0.3cm}
In this section, we present the computational advantages achieved by using {our optimal search schemes} in ASM-B (ASM performed completely in bidirectional FM-index).  %Section \ref{sec:famous} presents the computational advantages gained in ASM by using \viogray{FAMOUS}{the optimal search schemes}.
%\vspace*{-2ex}
}

{While our optimal search schemes can be found for any alphabet size and read length, we chose to concentrate on parameter values relevant to standard sequencing reads, e.g., Illumina reads.}
In Table \ref{tab:edges}, for a number of relevant parameter values, we have shown how the total number of edges using the optimal search schemes found by our MIP is reduced compared to  the unidirectional backtracking scheme. It can be seen that the reduction is between  $42\%$ and $49\%$. Also, for $K=2$ and $K=3$, the optimal search scheme with $P=K+2$ has fewer edges than the one with $P= K+1$. %\vspace{-2ex}%Varying the size of the partitions (data not shown) we can improve this by $\approx 3\%$. \todo{Chris: check}

\begin{table}[ht]
\setlength{\tabcolsep}{7pt}\renewcommand{\arraystretch}{1.1}
\caption{Total number of edges in the optimal search schemes found by our MIP for $K=1,2,3$ and $P=K+1$, $P=K+2$ and $P=K+3$ compared to full backtracking. The factor column shows the ratio of total number of edges in each scheme to that in backtracking. The optimal search schemes are listed in the Appendix in Table \ref{tab:schemes}.}\vspace{-4ex}

\begin{center}
\resizebox{\textwidth}{!}{
\begin{tabular}{c|l|r|r|r|r|r|r|r|r}
%\hline
\multirow{2}{*}{Distance}& \multicolumn{1}{c|}{\multirow{2}{*}{Search Scheme}} & \multicolumn{2}{c|}{$K = 1$} & \multicolumn{2}{c|}{$K = 2$} & \multicolumn{2}{c|}{$K = 3$} & \multicolumn{2}{c}{$K = 4$} \\
\cline{3-10}
&  & \multicolumn{1}{c|}{Edges} & Factor & \multicolumn{1}{c|}{Edges} & Factor & \multicolumn{1}{c|}{Edges} & Factor & \multicolumn{1}{c|}{Edges} & Factor \\
\hline
%\parbox[t]{2mm}{\multirow{3}{*}{\rotatebox[origin=c]{90}{Hamm}}} & BT	&	15,554 & 1.00	&	1,560,854 & 1.00	&	116,299,379 & 1.00 \\
{\multirow{4}{*}{Hamming }} & Backtracking	&	15,554 & 1.00	&	1,560,854 & 1.00	&	116,299,379 & 1.00 & 6,862,924,649 & 1.00 \\
&{Optimal ($P=K+1$)}     &	{8,004}	& 0.51 &	{892,769}	& 0.57 &	{67,888,328} & 0.58 & 4,064,852,156 & 0.59\\
& {Optimal ($P=K+2$)}     &	{8,922}	& 0.57  &	{854,303}	& 0.55 &	{65,116,676} & 0.56 & 3,916,700,994 & 0.57 \\
& {Optimal ($P=K+3$)}     &	{8,004}	& 0.51  &	{835,213}	& 0.54 &	{64,060,718} & 0.55 & 3,887,857,820 & 0.57 \\
\hline
%\parbox[t]{2mm}{\multirow{3}{*}{\rotatebox[origin=c]{90}{Edit}}} & BT	&	41,208 & 1.00	&	11,154,036 & 1.00	&	2,264,515,748 & 1.00 \\
{\multirow{4}{*}{Edit}} &Backtracking	&	41,208 & 1.00	&	11,154,036 & 1.00	&	2,264,515,748 & 1.00 & 367,846,294,116 & 1.00 \\
&{Optimal ($P=K+1$)}     &	{20,908} & 0.51	&	{6,315,779} & 0.57	&	{1,299,709,022} & 0.57 & 213,296,122,595 & 0.58 \\
& {Optimal ($P=K+2$)}   &	{23,356} & 0.57	&	{6,025,907} & 0.54	&	{1,246,126,103} & 0.55 & 205,509,484,572 & 0.56 \\
& {Optimal ($P=K+3$)}   &	{20,908} & 0.51	&	{5,892,667} & 0.53	&	{1,226,903,544} & 0.54 & 203,270,363,390 & 0.55 \\
\hline
\end{tabular}
}
\end{center}
\label{tab:edges}\vspace{-5ex}
\end{table}%\vspace{-5ex}

%Schemes:
%
%\begin{table}[ht]
%\begin{center}
%\begin{tabular}{|l|r|r|r|r|}
%\hline
%  & \multicolumn{1}{c|}{$e = 1$} & \multicolumn{1}{c|}{$e = 2$} & \multicolumn{1}{c|}{$e = 3$} \\
%\hline
%\textbf{SS K+1} & \makecell{(12, 00, 01) \\ (21, 01, 01)} & \makecell{(123, 002, 012) \\ (321, 000, 022) \\ (231, 011, 012)} & \makecell{(1234, 0003, 0233) \\ (2341, 0000, 1223) \\ (3421, 0022, 0033)} \\
%\textbf{SS K+2} & \makecell{(213, 001, 001) \\ (321, 000, 011)} & \makecell{(2134, 0011, 0022) \\ (3214, 0000, 0112) \\ (4321, 0002, 0122)} & \makecell{(12345, 00022, 00333) \\ (43215, 00000, 11223) \\ (54321, 00003, 02233)} \\
%\hline
%\end{tabular}
%\end{center}
%\end{table}

Although the reduction factors in total number of edges obtained by our optimal search schemes in Table \ref{tab:edges} are very significant in themselves, due to the stochastic nature of occurrence of errors in sequencing reads and occurrence of approximate matches in the reference genome, the real-case ASM speed-up factors achieved by these optimal search schemes compared to backtracking can be yet much more significant. To gain insight into this speed-up, we performed an experiment searching for \emph{all} approximate matches (for $K=1$, $2$, and $3$) of $100,000$ real Illumina reads of length $R=101$ (SRA accession number ERX1959065) in the human genome hg38 and compared the running time of {ASM-B performed with optimal search schemes obtained by our MIP for $P=K+1$ and $P=K+2$ to that of unidirectional backtracking for Hamming and edit distance}.  All of our tests were conducted on Debian GNU/Linux 7.1 with Intel Xeon E5-2667V2 CPUs at fixed frequency of 3.3 GHz to prevent dynamic overclocking effects. All data was stored on tmpfs, a virtual file system in main memory to prevent loading data just on demand during the search and thus affecting the speed of the search by I/O operations. All tools were run with a single thread to make the results comparable. %exclude effects of different speed-ups.
The results are shown in Table \ref{tab:runtime}. We can see that for both Hamming and edit distance, {employing} our optimal search schemes, is much faster than backtracking, verifying our expectation. The respective speed-ups for {$K=1$, $2$, and $3$ are $3.1$, $14.3$, and $35.2$} for Hamming distance, and {$4.1$, $11.1$, and $21.3$} for edit distance, much more significant than reduction in the total number of edges reported in Table \ref{tab:edges}.\vspace{-3ex}

%with options {\tt -a L 25 -i C,25,0} \todoki{Dear Knut and Chris, how do we control $K$ in Bowtie2? KnutL You cannot, hence we added BWA. FOr Bowtie2 we used the settings of Vroland. Alternatively we take Bowtie2 completely out of the paper.} to make it enumerate all alignments. Unfortunately, Bowtie2 did not terminate within 13 \todoki{Is 13 correct? Knut: Yes} hours with these settings, \todoki{correct? KNut: yes} neither did it terminate within 3 hours with the default configuration and option {\tt -a}.

{\setlength{\tabcolsep}{7pt}
\renewcommand{\arraystretch}{1.1}
\begin{table}[ht]
\caption{Running time comparison of searching all approximate matches of $100,000$ Illumina reads ($R=101$) using {optimal bidirectional scheme } with $P=K+1$ and $P=K+2$ versus backtracking for Hamming {and edit }distance.  The factor column is the speed-up ratio versus backtracking in each category.}

	\begin{center}
		\begin{tabular}{c|l|r|r|r|r|r|r}
			%\hline
			%\multirow{2}{*}{Distance}
			\parbox[t]{2mm}{\multirow{2}{*}{\rotatebox[origin=c]{90}{Dist.}}}& \multicolumn{1}{c|}{\multirow{2}{*}{Search Tool}} & \multicolumn{2}{c|}{$K = 1$} & \multicolumn{2}{c|}{$K = 2$} & \multicolumn{2}{c}{$K = 3$} \\
			\cline{3-8}
			&  & \multicolumn{1}{c|}{Time} & Factor & \multicolumn{1}{c|}{Time} & Factor & \multicolumn{1}{c|}{Time} & Factor \\
			\hline
			%\parbox[t]{2mm}{\multirow{3}{*}{\rotatebox[origin=c]{90}{Hamm}}} & BT	&	15,554 & 1.00	&	1,560,854 & 1.00	&	116,299,379 & 1.00 \\
			%{\multirow{3}{*}{Hamming }}
			\parbox[t]{2mm}{\multirow{3}{*}{\rotatebox[origin=c]{90}{Hamm.}}}& Backtracking	&	22.80s	&	1.00	&	269.24s	&	1.00	&	2417.06s	&	1.00 \\
			&\textbf{Optimal-scheme bidirect. ($P=K+1$)}     &	\textbf{7.73s}	&	\textbf{2.95}	&	\textbf{19.78s}	&	\textbf{13.61}	&	\textbf{74.62s}	&	\textbf{32.39} \\
			& \textbf{Optimal-scheme bidirect. ($P=K+2$)}    &	\textbf{7.39s}	&	\textbf{3.09}	&	\textbf{18.81s}	&	\textbf{14.31}	&	\textbf{68.69s}	&	\textbf{35.19} \\
            % & Bowtie 1    & 23.00s  &       0.99    &       68.00s  &       3.96 & 180.00 & 13.43    \\
			\hline
			%\parbox[t]{2mm}{\multirow{3}{*}{\rotatebox[origin=c]{90}{Edit}}} & BT	&	41,208 & 1.00	&	11,154,036 & 1.00	&	2,264,515,748 & 1.00 \\
			%{\multirow{3}{*}{Edit}}
			\parbox[t]{2mm}{\multirow{3}{*}{\rotatebox[origin=c]{90}{Edit}}} &Backtracking	&	43.59s	&	1.00	&	1245.70s	&	1.00	&	27889.40s	&	1.00 \\
			&\textbf{Optimal-scheme bidirect. ($P=K+1$)}    &	\textbf{11.21s}	&	\textbf{3.89}	&	\textbf{120.70s}	&	\textbf{10.32}	&	\textbf{1338.61s}	&	\textbf{20.83} \\
			& \textbf{Optimal-scheme bidirect. ($P=K+2$)}   &	\textbf{10.66s}	&	\textbf{4.09}	&	\textbf{112.23s}	&	\textbf{11.10}	&	\textbf{1307.23s}	&	\textbf{21.33} \\
			%& BWA	&	12.67s	&	3.44	&	143.32s	&	8.69	&	297.32s	&	93.80 \\
			%& Bwolo		&	20.79s	&	2.10	&	64.09s	&	19.44	&	208.19s	& 133.96 \\
			%& Yara   &	11.53s	&	3.78	&	70.56s	&	17.65	&	153.77s	& 181.37 \\
			\hline
		\end{tabular}\vspace*{-2ex}
	\end{center}

\label{tab:runtime}\vspace*{-5ex}
\end{table}
}
%\todoBen{Adding bowtie 1, which does not use dynamic programming, to table 2 may be a good idea for Hamming distance}
%\todoki{I removed $K=0$ from the tables, is that OK? Knut: Yes}
{We also compared the optimal search schemes with Hamming distance against Bowtie1~\cite{langmead2009ultrafast} by searching for all alignments with at most $K$ mismatches ({\tt -v <$K$> -a}). It turns out that our search schemes are significantly faster (Bowtie1 takes $23$, $68$ and $180$ seconds for $1$, $2$ and $3$ errors).} \vspace{-3ex}
%\todoki{I readded Bowtie1 results in the text. The original values are in Table2 hidden as a comment. Chris}

\section{Towards a full-fledged aligner}\vspace*{-2ex}
\label{sec:famous}
Due to the exponential complexity of ASM using FM-index in terms of $K$, the state-of-the-art aligners do not perform ASM completely in index but rather use a combination of search in the index and verification in text using dynamic programming (DP). Pure index-based search using standard backtracking is very slow for larger values of $K$. However, the drastic improvement, over standard backtracking, gained by using our optimal search schemes for search in an bidirectional index, as observed in Section \ref{sec:comp}, suggests a potential for significant improvement in the performance of read mappers that utilize index-based search. To acquire a sense of this potential, we decided to even challenge our optimal search schemes by using them in a pure index-based search and compare the results against the full-fledged state-of-the-art aligners that have the advantage of using a combination of index-based search and in-text verification using DP.  %and with the read mappers Bowtie2 \cite{Langmead2012a},

We performed our first set of comparisons with BWA \cite{Li:2009fi}, Yara \cite{Yara}, as well as an available implementation of the $01^*0$-filter scheme combined with dynamic programming (named Bwolo) \cite{Vroland:2016ca} in the all-mapping mode. We did these comparisons for the edit distance only as all these tools work for edit distance. As before, $100,000$ Illumina reads of length $R=101$ were aligned.
 BWA was run with the options {\tt -N -n <$K$>}, and Yara was run with the options {\tt -e <$K$> -s <$K$> -y full -t 1}. %\todoki{Does {\tt<errors>Knut: Yes} mean K everywhere?}
	We note that Bowtie2 \cite{Langmead2012a} is not designed with all-mapping in mind (for our data set, it did not terminate in 3 hours with default configuration and {\tt -a} option). Moreover, imposing an all-mapping with maximum $K$ errors in Bowtie2 in a way that its results are comparable to other tools is difficult. Bowtie2 settings used in \cite{Vroland:2016ca} do not enforce this, and nonetheless, led to a very long running time for our data set (did not terminate in 13 hours). Consequently, we did not use Bowtie2 in this study.\vspace*{-2ex}

\begin{remark}
We would like to note that while BWA and Yara are standard read mapping tools for Illumina reads, Bwolo is interesting in the context of our approach.
Vroland et al. \cite{Vroland:2016ca} presented this fast method for searching in an index by partitioning the read into $K+2$ pieces for $K$ errors and then exploiting the fact that the $K+2$ pieces must contain the mismatch pattern $01^*0$. Their method searches for an occurrence of the $01^*0$ mismatch pattern in the read and then verifies the remaining pieces of the read partially in the index (they use a unidirectional index) and partially in the text via dynamic programming. Interestingly, their $01^*0$ method can be formulated as a search scheme. For example, for $K=2$, and hence $P=4$, their method can be expressed as the following search scheme: $\mathcal{S}_{01^*0}=\{(4321,0000,0122),(3214,0000,0122),(2134,0000,0022)\}$.
%\todoki{Dear Knut and Chris, based on my understanding of Vroland's procedure $s_2$ in his procedure is also (3214,0000,0112) as he does not consider 0,1,1,0 mismatch pattern for 3,2,1,4 order. }\todo{No. I think that is correct. He does not consider 0,1,1,0 but 0,0. In the latter case he has to verify pieces 1 and 4 with 0 to 2 errors}
The  optimal search scheme found by our MIP for these parameters is \mbox{$\sopt=\{(4321,0002,0122),(3214,0000,0112),(2134,0011,0022)\}$}, which is quite similar, but has yet fewer edges. Indeed, we verified the practical superiority of $\sopt$ to $\mathcal{S}_{01^*0}$ by searching $100,000$ real Illumina reads in the human genome with $K=2$ Hamming and edit distance and noticed that $\sopt$ took less time than $\mathcal{S}_{01^*0}$ ($18.8$s vs. $22.0$s for Hamming, and $112$s vs. $160$s for edit distance, respectively). \qed
\end{remark}\vspace*{-3ex}

Back to our first set of comparisons, for $K=1$, BWA, Bwolo, and Yara, took $12.67$s, $20.79$s, and $11.53$s, respectively. This is while, according to Table \ref{tab:runtime}, the pure index-based ASM using our optimal schemes for $P=K+1=2$ and $P=K+2=3$, took $11.21$s, and $10.66$s, respectively, i.e., our optimal schemes are so effective that although the search is performed completely in index, the resulting times are better than these full-fledged state-of-the-art aligners, which use a combination of index search and in-text verification. Note that pure index-based search using standard backtracking takes $43.59$s which is much worse than all aforementioned times. For $K=2$, BWA, Bwolo, and Yara times took $143.32$s, $64.09$s, and $70.56$s, while pure index-based ASM using our optimal  schemes for $P=K+1=3$ and $P=K+2=4$, took $120.70$s, and $112.23$s, respectively, i.e., they performed better than BWA but worse than Bwolo and Yara. For $K=3$, the benefit of using in-text verification in full-fledged aligners catches up and all of them work faster than the pure index-based search using our optimal schemes.

We find these results for our optimal search schemes very impressive. To our knowledge, this is the first time that, for $K=1$ and $K=2$, ASM of reads of this size ($R=101$) performed completely in index has been reported to compete in running time with the best full-fledged aligners, which use combination of index search and in-text DP verification. This implies the power of our optimal search schemes.
We note that the results are even better in strata mode. We performed a second set of comparisons with Yara, this time in strata mode. The $0$-strata search means we first search the reads with $0$ errors, then search all the reads with no exact match, with $1$ error, and so on, until $K$ is reached. This strategy can be generalized to $s$-strata, where $s\le K$. This means that, for $b=0$ to $K-s$, for all reads with a $b$-error best match, we compute all occurrences with up to $b+s$ errors. We ran Yara in $1$-strata mode using {\tt -e <$K$> -s 0 -y full -t 1} and compared the time with the pure index-based search using our optimal schemes in the same mode. For $K=1$, $2$, and $3$, Yara took $3.57$s, $8.00$s, and $38.48$s, respectively, and the pure index-based search took $4.36$s, $8.57$s, and $53.74$s, respectively. That is for $K=1$, $2$, and $3$, the pure index-based search using our optimal search scheme  is about as fast as the full-fledged aligner Yara, which uses index search and in-text DP verification.

%For running FAMOUS in strata mode, the optimal search scheme for each value of $b$ is found using a modified version of our MIP in which a minimum number of errors is enforced.
%\todo{I think Chris did not do that but simply used the best search scheme. Since we did 1-strata, it did not matter. Chris? I used the original schemes with 0 as lower bound for K since we don't have the "best schemes" (like K+2 parts). And from my first tests, the speed-up is insignificant when increasing the lower bound. I think this becomes relevant when we write the read mapper and don't want to worry about duplicates.}
%We can see \viogray{that FAMOUS}{the results suggest using optimal search schemes } is about as fast as Yara for $K=1$ and $2$, and slightly worse for $K=3$,  which is consistent with the results of our first experiment. The difference in the $1$-strata mode is that we only have to map about $7\%$
%\todo{Chris: it's exactly 7,187}
%of the $100,000$ reads with $3$ errors, which does not affect the overall running time as heavily as in \viogray{the first}{previous} experiment. \vspace*{2ex}

Of course, we did not expect to be able to outperform full-fledged aligners for larger values of $K$, because for larger $K$ and this read length, verification in index is too costly, especially if the number of successful verifications is low (in our case almost all reads occur only once in the genome).
Although Vroland et al. \cite{Vroland:2016ca} raised the option of using a bidirectional index for verification, they only used in-text DP verification for the last pieces as they had only a unidirectional index at hand. Nevertheless, interestingly, for their data set (40bp, exactly 3 errors), pure index-based search using our optimal search schemes outperforms Bwolo by a factor of almost $1.5$ (data not shown), so the read length matters. {Although the optimal scheme found by our MIP is superior to the $01^*0$ scheme of Vroland et al. \cite{Vroland:2016ca} for search in the index, for a larger $K$, one has to take into account the number of remaining verifications versus the number of edges in the trie for the remaining pieces. If that ratio is low, it does not pay off to verify in the index instead of verification in the text as our comparisons showed. }

Our observations in this section about the power of our optimal search schemes suggest that a full-fledged aligner that employs an intelligent combination of search in the bidirectional FM-index using our optimal search schemes and in-text verification using DP can outperform today's best approximate read mappers. The development of such an aligner, called FAMOUS (Fast Approximate string Matching using OptimUm search Schemes) is ongoing in our group as future work.\vspace*{-2.5ex}

\section{Conclusions}\vspace*{-2ex}%\todoki{I am done revising up to here}.
\label{sec:con}
{In this paper, we contributed to the approximate string matching research as follows:}\vspace*{-2ex}
\begin{enumerate}
\item {We proposed,} for the first time, a method to solve the optimal search scheme problem for ASM-B for Hamming distance (using a MIP formulation).
\item %\viogray{In addition, we presented FAMOUS (Fast Approximate string Matching using OptimUm search Schemes), a bidirectional search (for Hamming and edit distance) implemented in SeqAn \cite{Reinert:2017gm} that performs the search based on optimal search schemes obtained from our MIP and showed that FAMOUS is up to $35$ times faster than standard backtracking. }
{We demonstrated that our MIP approach can solve the optimum search scheme problem to optimality in a reasonable amount of time for input parameters of considerable size, and enjoys  very quick convergence to optimal or near-optimal solutions for input parameters of larger size.

\item We showed that approximate search in a bidirectional FM-index can be performed significantly faster if the optimal  schemes obtained from our MIP are used. This was demonstrated based on number of edges in the search tries as well as actual running time of in-index search on real Illumina reads (up to $35$ times faster than standard backtracking for 3 errors). We also showed that although our MIP solutions are for Hamming distance they perform equally well for edit distance.
	}
\item We showcased the power of our optimal search schemes by demonstrating that for $K=1$ and $2$ errors, approximate string matching of reads of size $R=101$ performed completely in index compete in running time with the best full-fledged aligners, which benefit from combining search in index with  in-text dynamic programming verification. This suggests that a full-fledged aligner that intelligently combines search in bidirectional index using our optimal search schemes with in-text verification using DP can outperform today's best approximate aligners. The development of such an aligner, called FAMOUS is ongoing as our future work.\vspace*{-1ex}
\end{enumerate}
{Moreover, our approach in this research has raised some interesting open problems}:\vspace*{-2ex}
\begin{enumerate}\parskip=0.2ex
%\begin{description}
	\item Our computational experiments in  Section \ref{sec:mipcomp} showed that our current MIP has two attractive properties: the early solutions it finds are optimal or near-optimal, and its optimal search scheme is insensitive to the value of $R$ (we ask: \emph{``is this insensitivity to $R$ a theoretically provable fact?''}). This makes our current MIP quite powerful in practice because, even if all input parameters $K$, $R$, $P$, $\os$ are quite large, we can run the MIP for a short time with a much smaller $R$ to get a solution that is most probably optimal or near-optimal for the original problem. Nevertheless, solving the MIP completely to ascertain optimality is of great interest and currently consumes considerable computational resources for large instances, especially when $\os>5$, $K>4$, $P>6$, $R>100$. We ask \emph{``can the solution time be improved by introducing other MIP formulations, or strengthening the current formulation  using strong  cutting planes or further elimination of symmetric solutions?''}

\item
Our current MIP is restricted to optimizing over equal-size pieces, with $P$ and $\os$ given as part of the input. We ask \emph{``are there (MIP) approaches to optimize over the number of pieces in the partition, and/or unequal piece lengths, and/or with no upper bound on the number of searches?''}

\item
Our optimal search scheme, $\sopt$, for the example of Lam et al. \cite{LamLiTam09} covers every possible mismatch pattern only once, ensuring that no duplicate computational effort is spent on the same mismatch pattern. We, however, believe that this is not always the case for every optimal search scheme. We ask \emph{``how would enforcing this requirement on the solution of MIP affect its solution time and the performance of optimal search schemes it obtains?''}

 \item
We demonstrated that solutions found by our MIP, although for Hamming distance, perform very well for the edit distance as well. We ask \emph{``are there (MIP) approaches to find the actual optimal search scheme for the edit distance?''}

\item
We demonstrated that the verification of few occurrences with high errors in the index is worse than in-text DP verification. We ask \emph{``what is the best point to stop verification in the index and start verifying in the text instead?.''} This can be individually decided for each pattern. We are addressing this problem in the ongoing development of our upcoming full-fledged aligner FAMOUS.\vspace*{-3ex}

%\end{description}
\end{enumerate}

% add open questions

%%%%%%
\section*{Acknowledgments}\small\vspace*{-2ex}
We acknowledge Texas A\&M University High Performance Research Computing (HPRC) for providing resources to perform parts of computational experiments. The second author also acknowledges the support of the International Max-Planck Research School for Computational Biology and Scientific Computing (IMPRS-CBSC).\vspace*{-4ex}

\bibliographystyle{splncs_srt}
\bibliography{OptimalSearchPaper}

\section*{Appendix}\vspace*{-4ex}

\begin{figure}[H]
\noindent\begin{minipage}{3.4in}
\centering
\includegraphics[width=3in]{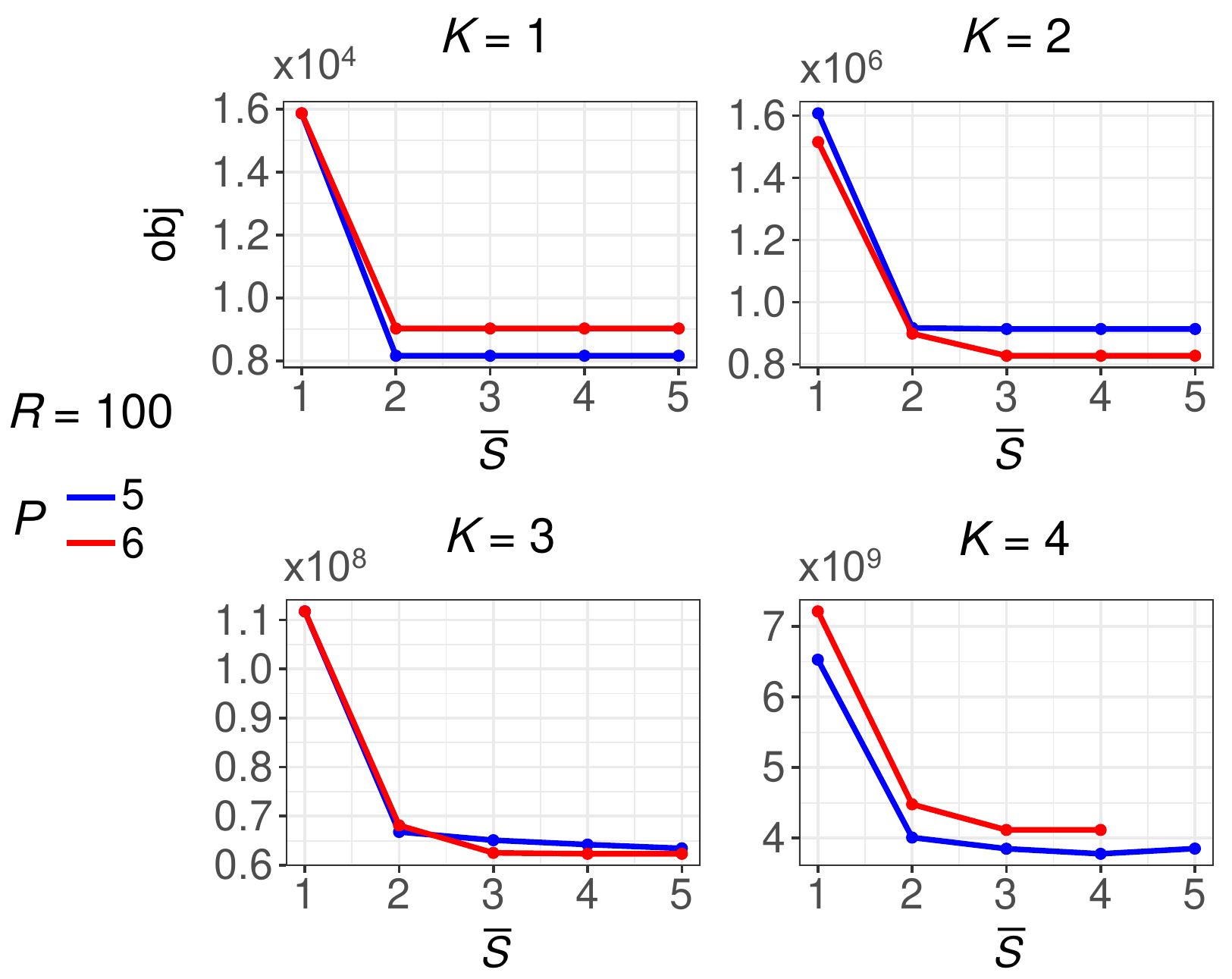}

\small\quad(a)
\end{minipage}%\hspace{0.2in}
\begin{minipage}{3.4in}
\centering
	\includegraphics[width=3in]{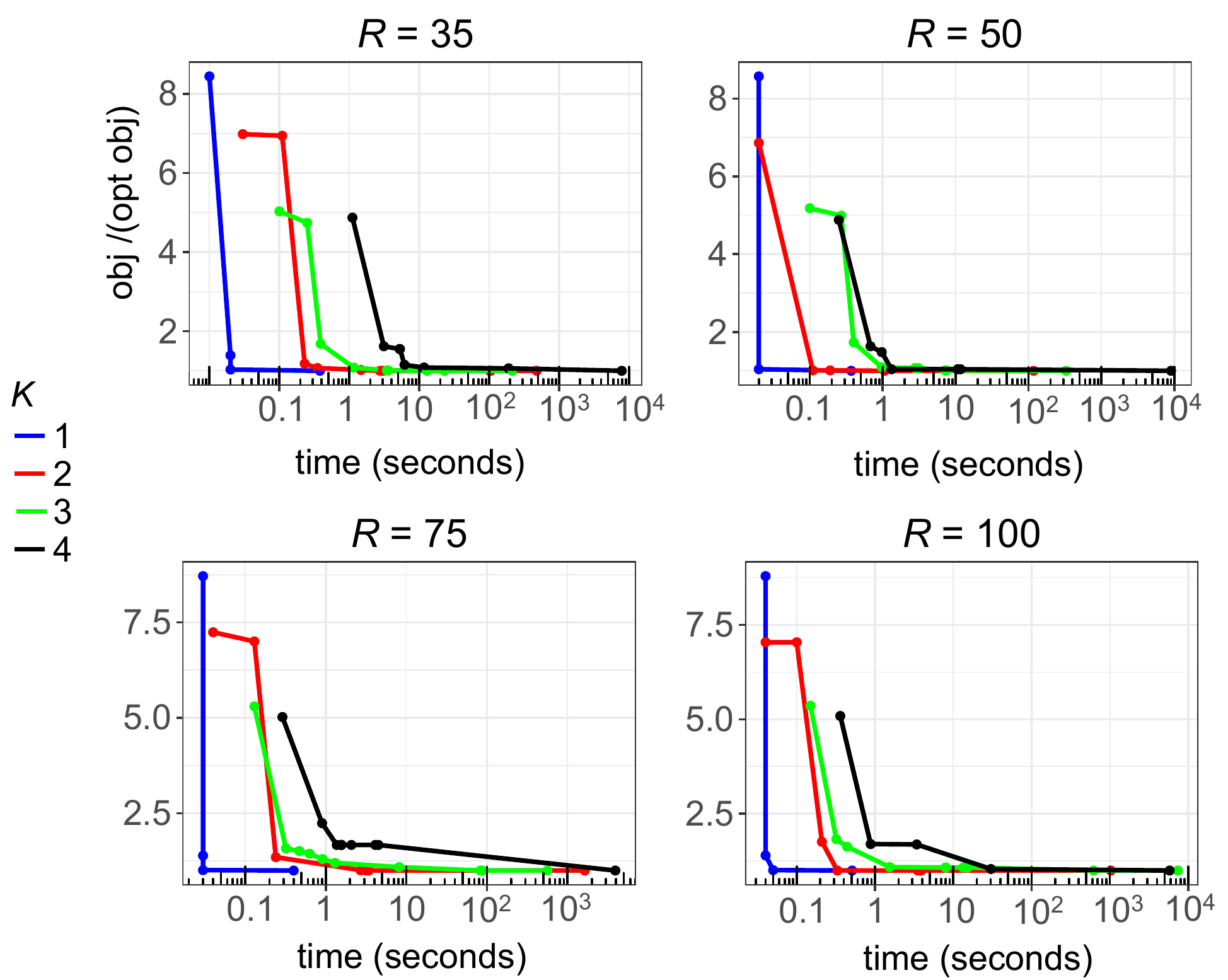}

\small\quad(b)
\end{minipage}%\hspace{0.2in}
\vspace*{-2ex}

\caption{{\bf (a)}. Sensitivity of optimal objective value to parameters $R$, $K$, $\os$, and $P$. For some cases due to memory overflow, there is no data point.
{\bf (b)}  Rapid convergence of feasible solutions to the optimal solution.
 \vspace*{-0.1in}
} \label{fig:graph}
\end{figure}

\begin{table}[H]
\setlength{\tabcolsep}{7pt}\renewcommand{\arraystretch}{1.1}
\caption{Search schemes found by our MIP for $K=1,2,3$ and $P=K+1$, $P=K+2$ and $P=K+3$ used for experiments in Tables \ref{tab:edges} and \ref{tab:runtime}. The schemes for $K=1$ and $2$ are optimal schemes with $\os=5$. To control the running time of MIP, the schemes for $K=3$ and $4$ are best solutions found by running the MIP for 2 hours with $\os=3$. These schemes are most probably optimal for $\os=3$.}

\begin{center}
\resizebox{\textwidth}{!}{
\begin{tabular}{l|c|c|c|c}
%\hline
& $K = 1$ & $K = 2$ & $K = 3$ & $K = 4$\\
\hline
%\parbox[t]{2mm}{\multirow{3}{*}{\rotatebox[origin=c]{90}{Hamm}}} & BT	&	15,554 & 1.00	&	1,560,854 & 1.00	&	116,299,379 & 1.00 \\
{Optimal ($P=K+1$)}     &
\makecell{$(12,00,01)$ \\ $(21,01,01)$} &
\makecell{$(123,002,012)$ \\ $(321,000,022)$ \\ $(231,011,012)$} &
\makecell{$(1234,0003,0233)$ \\ $(2341,0000,1223)$ \\ $(3421,0022,0033)$} &
\makecell{$(12345,00004,03344)$ \\ $(23451,00000,22334)$ \\ $(54321,00033,00444)$} \\
\hline
{Optimal ($P=K+2$)}     &
\makecell{$(123,001,001)$ \\ $(321,000,011)$} &
\makecell{$(2134,0011,0022)$ \\ $(3214,0000,0112)$ \\ $(4321,0002,0122)$} &
\makecell{$(12345,00022,00333)$ \\ $(43215,00000,11223)$ \\ $(54321,00003,02233)$} &
\makecell{$(123456,000004,033344)$ \\ $(234561,000000,222334)$ \\ $(654321,000033,004444)$} \\
\hline
{Optimal ($P=K+3$)}     &
\makecell{$(1234,0000,0011)$ \\ $(4321,0001,0011)$} &
\makecell{$(21345,00011,00222)$ \\ $(43215,00000,00112)$ \\ $(54321,00002,01122)$} &
\makecell{$(123456,000003,022233)$ \\ $(234561,000000,111223)$ \\ $(654321,000022,003333)$} &
\makecell{$(1234567,0111111,3333334)$ \\ $(1234567,0000000,0044444)$ \\ $(7654321,0000004,0333344)$} \\
\hline
\end{tabular}
}
\label{tab:schemes}
\end{center}
\end{table}\vspace{2ex}
%\newpage
%
%
%\section*{Appendix}
%In the appendix we give
\end{document}